\documentclass{achemso}

\usepackage[hidelinks]{hyperref}
\usepackage{graphicx}
\usepackage{color}
\usepackage{bm}
\usepackage{siunitx}
\usepackage{filecontents}
\usepackage[utf8]{inputenc}
\usepackage{amsmath}

\title{Thermal origin of light emission in non-resonant and resonant tunnel junctions}

\author{Christian Ott}
\affiliation{Department Physik, Friedrich-Alexander-Universit\"at Erlangen-N\"urnberg (FAU), Staudtstr. 7, D-91058 Erlangen, Germany}
\author{Stephan Götzinger}
\affiliation{Department Physik, Friedrich-Alexander-Universit\"at Erlangen-N\"urnberg (FAU), Staudtstr. 7, D-91058 Erlangen, Germany}
\alsoaffiliation{Max Planck Institute for the Science of Light, Staudtstr. 2, D-91058 Erlangen, Germany}
\alsoaffiliation{Graduate School in Advanced Optical Technologies (SAOT), FAU, D-91052 Erlangen, Germany}
\author{Heiko B. Weber}
\affiliation{Department Physik, Friedrich-Alexander-Universit\"at Erlangen-N\"urnberg (FAU), Staudtstr. 7, D-91058 Erlangen, Germany}
\email{heiko.weber@fau.de}

\begin{document}

\begin{abstract}
Electron tunneling is associated with light emission. In order to elucidate its generating mechanism, we provide a novel experimental ansatz that employs fixed-distance epitaxial graphene as metallic electrodes. In contrast to previous experiments, this permits an unobscured light spread from the tunnel junction, enabling both a reliable calibration of the visible to infrared emission spectrum and a detailed analysis of the dependence of the parameters involved. 
In an open, non-resonant geometry, the emitted light is perfectly characterized by a Planck spectrum. In an electromagnetically resonant environment, resonant radiation is added to the thermal spectrum, both being strictly proportional in intensity. In full agreement with a simple heat conduction model, we provide evidence that in both cases the light emission stems from a hot electronic subsystem in interaction with its linear electromagnetic environment. These very clear results should resolve any ambiguity about the mechanism of light emission in nano contacts.
\end{abstract}


It is known since the early days of scanning tunneling microscopy, and even before, that electron tunneling from a metallic tip to a metallic surface is associated with light emission\cite{bor65, lam76, gim88}. The explanation is commonly referring to the discrete nature of charge:  the tunneling current consists of a sequence of single electron bursts obeying Poisson statistics, which not only generates shot noise, but is also capable of triggering electromagnetic modes, in particular resonant surface plasmons\cite{pet17, sch09, ker15, lam76, kaa15, xu14, var16, nam18, sch10}. It is, however, not easy to prove this narrative. Closest to an experimental confirmation are a subtle substructure of light intensity in the overbias regime\cite{pet17} and a correlation to shot noise\cite{sch10}. Both will be critically revisited below.

Obviously, STM geometries, where this effect is very successfully exploited to even map the internal structure of a molecule via light emission\cite{zha17}, are not well suited to clarify the physical mechanism of light generation. Light that is created in the tunnel junction is detected in the far field, i.e. the evolution of the signal is affected by nano-optical phenomena (in a rather difficult geometry), including antenna-like or plasmon-like electromagnetic resonances\cite{est12}. A minute change of the tip distance (assume voltage being fixed) changes the current, the tip-plasmon coupling, the impedance that rules the evolution towards the far field, effects that can barely be disentangled\cite{kuh17}. In recent experiments on planar metallic nanojunctions, predominantly blackbody-like light emission (in a small spectral range) is reported, plus nonlinear resonant features that were linked to quantized plasmon excitations\cite{bur15}. This underscores concerns whether the underlying physics is well understood.

In order to make a complementary experimental ansatz, we opted for a completely different choice of materials and geometry: we used epitaxial graphene on silicon carbide. The result is essentially a metal-metal tunnel junction with barely detectable plasmonic effects in the respective range\cite{PeresPlasmonics}. Moreover it provides advantageous features: it is planar, unobscured, open-lying and nearly fully transparent. This allows for unprecedented and accurate spectral calibration of the emitted light, with a consistent data set from \SIrange{550}{1600}{\nano\meter}. With the obtained data, we present convincing experimental evidence that the emitted light stems from a hot electronic subsystem.

When electrons tunnel from a metal to a metal driven by a voltage $V$, they enter the opposite electrode as hot injected electrons. Their energy/heat is then redistributed within the electronic system via electron-electron scattering. In graphene, this electronic thermalization appears within \SIrange{20}{30}{\femto\second}, similar timescales are observed in common metals\cite{kno98}. On much longer timescales, in the picosecond range, electron phonon cooling sets in, which is transferring the heat out of the electronic system\cite{els87, gie13}.
Hence, within picoseconds the Joule heat is conserved in the electronic system and spreads outwards obeying the heat diffusion equation. In the stationary state, this means that at the injection spot the electron system is overheated in a two-fold sense: (i) there is a spatial decay of effective temperature $T_\mathrm{el,hot}(r)$ towards equilibrium in the electronic system and (ii) this temperature detaches thermally from the underlying vibrational system.

The simplest description of the stationary case can be made by consulting the phenomenological Wiedemann-Franz-law\cite{fra53}. We assume that there is a hot spot  with radius $r_1$ at the junction where the heat is injected into the electronic system. Further there is electronic cooling towards a thermally equilibrated bath with temperature $T_B$ at a larger radius $r_2$, where the heat is transferred to the lattice. Because in this inner region, the transport is purely electronic, the thermal conductivity $\lambda = \mathcal{L} \sigma T$ (The Lorenz number $\mathcal{L}$ equals  $\pi^2 k_B^2/(3e^2)$ in a Drude description) determines the thermal gradient, $\sigma$ is the electrical conductivity. As a result, at the tunnel junction, the electronic temperature is determined by 
\begin{equation}
T_\mathrm{el,hot} = \sqrt{\frac{P_\mathrm{el}}{2\pi\mathcal{L}\sigma} \log\left(\frac{r_2}{r_1}\right) + T_B^2}
\label{eq:TJ}
\end{equation}
for two dimensional heat spread. Note that in the high temperature limit $T_{el, hot} \propto \sqrt{P_{el}}$. When we insert realistic numbers for a graphene tunnel junction, temperatures of a few thousand Kelvin can be reached. Such a hot electron system, however, emits light with a thermal spectrum. While we do not expect quantitative predictive power from this model, we will see that despite its simplicity (as compared to the models used in Refs.\,\cite{bur15, xu14, kaa15}) this model gives a convincing description of nearly all phenomena.

The graphene nanojunctions are fabricated in a stepwise process. Starting with a lithographically defined graphene constriction, a controlled electroburning protocol\cite{ull15} is performed, which leads via a graphene point contact (GPC) regime close to quantum conductance in which presumably a single carbon bond forms the junction, to finally a graphene tunnel contact (GTC). 

First, we discuss the tunnel regime (GTC, Fig.\,1a), in which the tunnel resistance is at least \SI{100}{\kilo\ohm}. The IV characteristics display nonlinearities, from which we infer that the graphene lattice is interrupted. The tunnel pathway and the termination of the graphene sheet at the narrowest point is unknown. We focus on the observation that light is emitted from a narrow spot of approximately one micrometer size, i.\,e. at or below the optical resolution limit, when the voltage exceeds $\approx$\SI{1.5}{\volt}. We performed a spectral analysis of the emitted light in the visible and near-infrared regime with two different detectors. A careful spectral calibration was performed by replacing the sample holder by a macroscopic heated cavity with controlled temperature. Its blackbody radiation served as standard to calibrate the full optical pathway and the detectors. Reliable results were obtained above \SI{550}{\nano\meter}, at lower wavelengths the calibration provides inaccuracies and is not further considered. Fig.\,1c displays the associated corrected emission spectra. The underlying raw and corrected data are available online and briefly discussed in the Supplementary Information. For each voltage the spectrum is broad and resembles a Planck spectrum, however with deviations. We performed a Planck fit to the infrared part of the data (\SIrange{1000}{1600}{\nano\meter}), for which the temperature $T_\mathrm{Planck}$ and a proportionality constant are the only free parameters\cite{pla01}. It matches well and results in temperatures well beyond \SI{2000}{\kelvin}, see Fig.\,1c. 
We assign this broad signal to thermal radiation, which will become much more evident when going one step back in our fabrication scheme.
	
Before the electroburning process reaches the tunnel regime an important threshold in conductance is the GPC regime in which the junction conductance is close to quantum conductance. In the description of charge transport through nanojunctions, it plays a particular role because here the transmission, the shot noise amplitude and the light emission should be strongly dependent on details\cite{kum96, rez95, sch10}. In graphene, the contact is then presumably formed by a single or few C-C bonds\cite{sta08, sad15}, c.\,f. Fig.\,1b. This regime is particularly stable: when applying a voltage to the GPC, under ambient conditions its light emission can be seen with the bare eye up to a bias voltage of approximately \SI{3}{\volt}, beyond which the device fails. For a sample mounted in cryogenic vacuum, the voltage range can be significantly extended up to \SI{8}{\volt}, sometimes even \SI{10}{\volt}. The data have strong similarity to the GTC case, in particular the curves increase monotonously with bias. Here however, for any voltage, we find excellent match of the full emission spectra to Planck’s formula (see Fig.\,1d), reaching \SI{2300}{\kelvin} at \SI{8}{\volt}. Notably, despite both graphene and SiC are extremely robust materials, at these temperatures SiC would thermally decompose\cite{her11, emt09}. Even in the presence of air, $T_\mathrm{Planck}$ as high as \SI{1600}{\kelvin} was observed, a temperature at which the graphene electrodes would simply burn away. We conclude from these purely experimental findings that in GPCs there are subsystems at different temperatures. As dissipation starts with an electronic process which is subsequently cooled by electron-phonon relaxation processes\cite{gie13, ruz10}, we can assign the light emission with its $T_\mathrm{Planck}$ to the electronic system, whereas in particular the carbon lattice must be significantly colder, below the degradation threshold. Given the similarity and the continuous evolution of the spectra it can be concluded that also in the GTC regime similar physics rule, i.\,e. a hot electron subsystem determines the light emission.

Within the picture of an overheated electron temperature at the junction, the differences between GTC and GPC can readily be understood: In the GTC case, a large fraction of the applied voltage drops along the tunnel distance, i.\,e. within roughly \SI{1}{\nano\meter}. The result is a hot electron subsystem, the diameter of which can be estimated to be less than the electron-phonon scattering length scale, in accordance with the resolution limited spot. When, however, the junction is in the GPC case, the point contact resistance, the spreading resistance and the electric leads are all in the \SI{10}{\kilo\ohm} range, hence, the sharp voltage drop at the nanojunction is strongly diminished. This goes along with a reduced electronic temperature. Another qualitative confirmation can be achieved by changing the material (quasi-freestanding bilayer graphene on SiC is used, see S.I. section 6) such that the electron-phonon scattering rate is lowered. There, much higher electronic temperatures can be achieved with the same applied bias, extending light emission into the so called overbias regime\cite{pet17}, a category that turns out to be irrelevant. Because in the GPC case the spectral information is particularly clear and the stability is outstanding, it is well suited to study the underlying mechanisms in detail.

First, the concept of the overheated electronic subsystem is investigated. According to Eq.\,\eqref{eq:TJ}, $T_\mathrm{el,hot}(P_\mathrm{el})$ is essentially linear above a certain threshold. The large spectral range gives immediate access to the Planck temperature. Indeed, the data follow a linear relation, however with a kink that presumably indicates that an additional cooling mechanism comes into play, see Fig.\,2a. We suspect that a fraction of the injected electrons above the work function of graphene leave the material and thus reduce $P_\mathrm{el}$. In a next step, we investigate the power dependence of the emitted light. Because the junction distance is fixed and the material system SiC/graphene withstands very high electric fields, we have access to an unusually broad range. Fig.\,2b presents the intensity measured via avalanche photo detectors, i.\,e. counting photons up to approximately \SI{900}{\nano\meter}. In this representation, the intensity appears as a straight line over six orders of magnitude in intensity, which emphasizes that only one, clearly defined mechanism rules over the entire range. This behavior can readily be understood: when inserting Eq.\,\eqref{eq:TJ} into Planck's law, such a behavior follows immediately (see S.I. section 4). Hence, these high-quality spectra and voltage dependencies prove that the origin of light emission is thermal and the simple heating model is presumably the simplest valid description.

We now readdress the tunnel regime (GTC). As we can determine the Planck temperature from the infrared part of the spectrum, we subtract the Planck part from the experimental data and obtain a resonance-like contribution, centered around \SI{720}{\nano\meter}, see Fig.\,2c, which appeared as shoulder-like anomaly in Fig.\,1c. Its intensity increases with increasing voltage without a spectral shift. Even the peak shape is maintained but slightly differs from sample to sample like a fingerprint. This feature, being absent in GPCs, is sensitive to atomistic details of the tunnel junction and offers opportunities: we understand it as a local resonant electromagnetic environment of the tunnel process \cite{woj19, nat19}. This allows for an immediate comparison to STM experiments, where the interaction of the tunneling electrons and the resonant (plasmonic) modes are the essence of the physical concept\cite{xu14, kaa15, pet17}. We quantify the intensity of our resonant feature as a function of voltage (in a spectral window from \SIrange{690}{750}{\nano\meter}) and compare it with the spectral weight of the underlying thermal radiation in the very same window (see Fig.\,2d). It turns out that both are strictly proportional, without any threshold behavior or other nonlinearity. This observation is in contrast to\cite{bur15}, where in a faint signal a threshold and a strong nonlinearity has been suspected, which was assigned to quantum excitations of plasmons. We can only speculate on the origin of this electromagnetic resonance: we assume it is a graphene plasmon pinned to the sharply defined tunnel junction, further it might be supported by the presence of quasi-periodic stacking fault patterns that are ubiquitous in epitaxial graphene on SiC\cite{but14}. Note that the electromagnetic impedance of the environment remains entirely constant in our experiments, as no tip is moved. We conclude from our experiments that in the presence of a resonant electromagnetic environment, the thermal radiation is resonantly and linearly enhanced. The physics we observe is entirely unrelated to the granular nature of charge.

As we could demonstrate that both in non-resonant conditions and in resonant conditions it is a hot electronic subsystem that feeds both the Planck spectrum and the resonant spectrum, consequences of this findings are briefly discussed. Hot electrons with energies above the work function will leave the material and populate the open space surrounding the tunnel junction. As this open space is threaded with strong static electric field, and in particular in STM geometries heavily confined, electrons are hindered from escaping. We propose as a result a stationary charge density of hot and constantly heated electrons (i.e. a one-component plasma) in and the tunnel pathway. Not only it appears plausible that this mechanism suppresses electromagnetic emission of the granular tunnel current, it may also couple to the light field, which is an interesting question that deserves further research.

For graphene junctions, it becomes clear that nearly all phenomena can be well understood by the simple heat spread model both in the non-resonant GPC and the resonant GTC case. When comparing these findings to STM experiments\cite{gim88, pet17, bea14} that were interpreted in terms of electromagnetic trigger, one may ask why the very similar phenomenology (see S.I. section 5) should root in a different underlying mechanism. The effect of overbias emission has no special meaning in the thermal picture \cite{nat19}. Whether or not a tiny substructure at multiples of the applied voltage that appears in STM \cite{pet17}, but not in our experiments, is indicative of the generating mechanism of the main signal, is unclear. Another indication, an apparent correlation to shot noise\cite{sch10} should not be overinterpreted because it is reported that upon contact thermal noise exceeds the shot noise by far (noise temperature of \SI{2000}{\kelvin}) which rather confirms the thermal picture. Hence, while the thermal picture is simple and experimental confirmed, we find no convincing argument that evidences the electromagnetic excitation picture.

To conclude, we measured voltage dependent spectra of light emission when electrons tunnel from graphene to graphene. The extensive data show unambiguously that the light source is essentially a hot electronic subsystem, far above the lattice temperature, in agreement with a simple heat-spread model. Sharp tunnel gaps provide further an electromagnetic resonance, which reveals that the thermal light can be spectrally shaped: hot electrons, thermal radiation and resonant excitation are rigidly connected phenomena. According to our experiments, in tunnel junctions with point-like injection, the luminescence is driven by Joule heating of the electronic system.

\begin{itemize}
  \item This work was supported by the Deutsche Forschungsgemeinschaft (DFG), Project No. 182849149 (SFB 953). We acknowledge useful discussions with M. A. Schneider, R. Berndt, and U. Peschl.
  \item The authors declare that they have no competing financial interests.
  \item The data are available in an open access repository under DOI:10.22000/324.
	\item Supplementary information are provided online.
\end{itemize}

\bibliography{nanoletters_bibliography}

\begin{figure}[t]
  \begin{center}
    \includegraphics[width=0.7\textwidth]{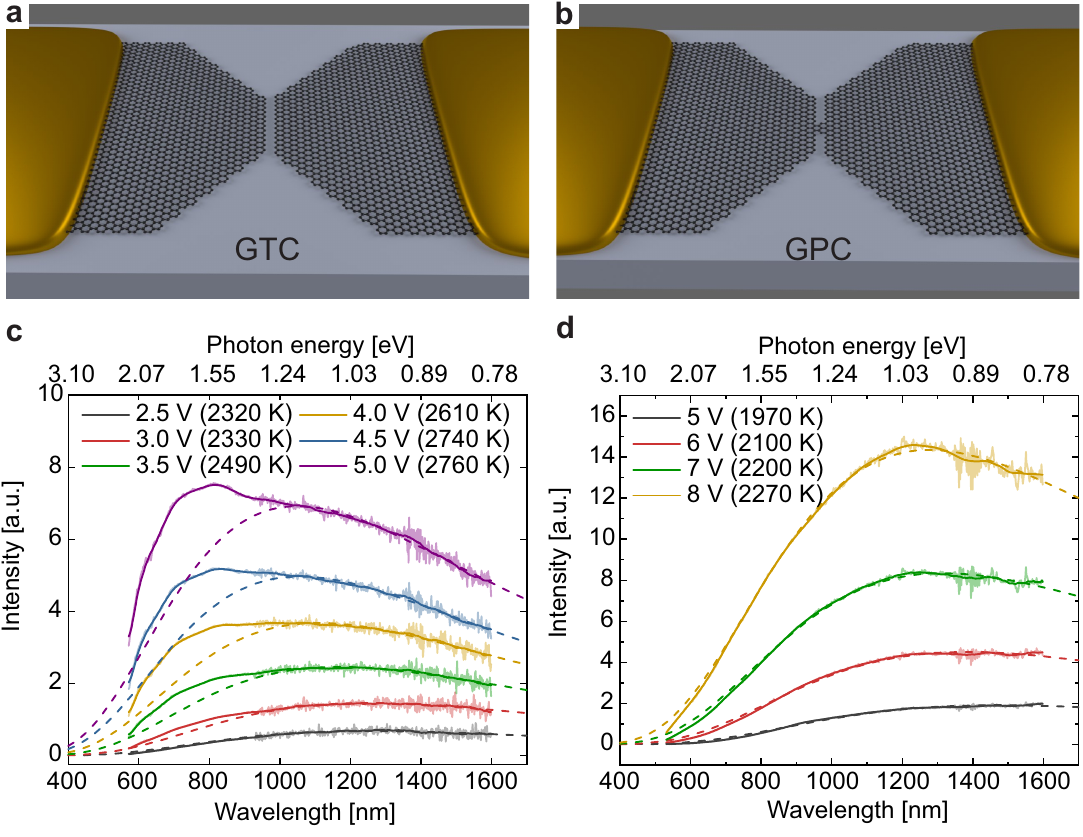}
    \caption{(a, b) Sketch of a graphene tunnel contact (GTC) and a graphene point contact (GPC), in which presumably a single carbon-carbon bond forms the contact. (c) For the GTC, the light emission is plotted as a function of the detection wavelength for various bias voltages. The spectra resemble Planck spectra, however with added spectral weight around \SI{720}{\nano\meter} that is further analyzed in Fig.\,2c. Dashed lines indicate fits of Planck's law to the IR data, the obtained temperatures are given in the legend. (d) For GPC, spectra are in good agreement with Planck's law displaying high electronic temperatures well beyond the damage threshold of the material.}
  \end{center}
\end{figure}

\begin{figure}[t]
  \begin{center}
    \includegraphics[width=\textwidth]{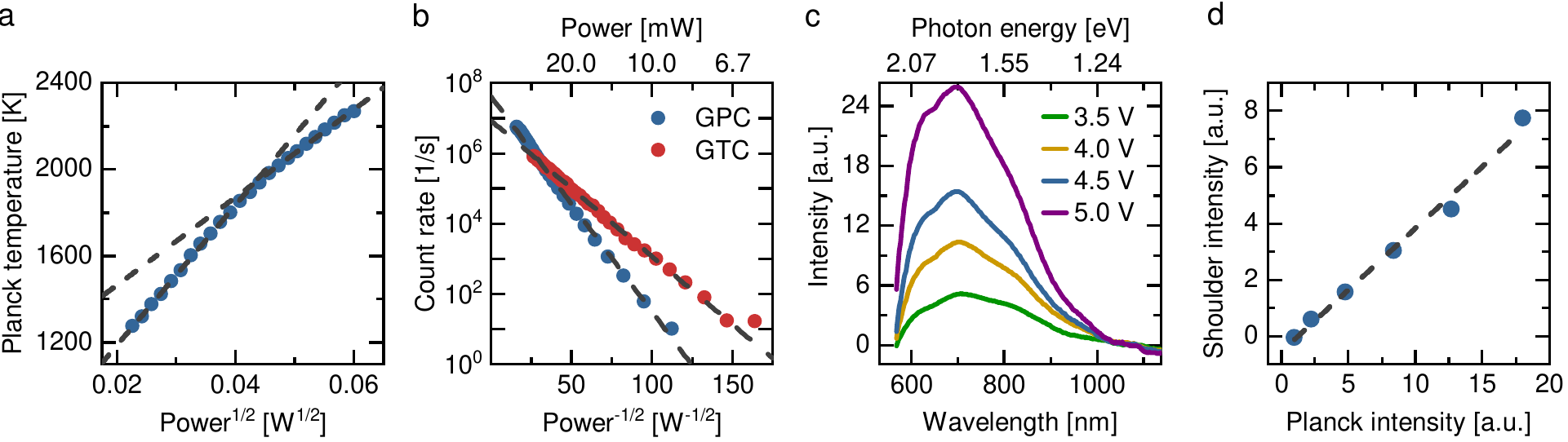}
    \caption{(a) Planck temperature as derived from the Planck spectra as a function of $\sqrt{P_\mathrm{el}}$ of a GPC sample. Two strictly linear regions can be identified in full accordance to Eq.\,\ref{eq:TJ}, the kink may indicate a crossover in cooling mechanisms. (b) Intensity dependencies on $P_\mathrm{el}^{-1/2}$ (GPC and GTC samples), which on the logarithmic scale appears as a straight line over six orders of magnitude (the constant dark count rate has been subtracted). (c) Analysis of the shoulder-like feature in Fig.\,1c, which appears as a self-similar resonance. (d) The amplitude of the latter scales strictly linear with the amplitude of the underlying blackbody radiation. The simplicity of all parameter dependencies (dashed straight lines are guides to the eye) is remarkable and underscores the validity of the model.}
  \end{center}
\end{figure}

\end{document}